\documentclass{llncs}

\usepackage{enumerate}
\usepackage{enumitem} 
\usepackage[utf8]{inputenc} 
\usepackage{graphicx}

\begin{document}

\title{Global Network Cooperation Catalysed by\\a Small Prosocial Migrant Clique} 

\author{Steve Miller\inst{1} \and Joshua Knowles\inst{2}}

\institute{School of Computer Science, University of Manchester, Manchester, UK
\and School of Computer Science, University of Birmingham, Birmingham, UK
\email{stevemiller.gm@gmail.com}}

\maketitle

\begin{abstract}
Much research has been carried out to understand the emergence of cooperation in simulated social networks of competing individuals.  Such research typically implements a population as a single connected network. Here we adopt a more realistic premise; namely that populations consist of multiple networks, whose members migrate from one to another.  Specifically, we isolate the key elements of the scenario where a minority of members from a cooperative network migrate to a network populated by defectors. Using the public goods game to model group-wise cooperation, we find that under certain circumstances, the concerted actions of a trivial number of such migrants will catalyse widespread behavioural change throughout an entire population. Such results support a wider argument: that the general presence of some form of disruption contributes to the emergence of cooperation in social networks, and consequently that simpler models may encode a determinism that precludes the emergence of cooperation.

\keywords{Evolution of cooperation $\cdot$  Evolutionary game theory $\cdot$ Public goods game $\cdot$ Complex networks}

\end{abstract}

\section{Introduction}

A considerable amount of scientific work has been undertaken to explain the apparently paradoxical existence of cooperative behaviour in a world defined by the competitive basis of natural selection~\cite{axelrod_evolution_1981}. The question of how cooperation may emerge within a competitive environment is, by definition, predicated on cooperation being originally absent from the population. On such a basis, the original appearance of cooperation occurs as a random event, more specifically, a mutant behaviour in (rare) individual(s). We then consider whether such a mutation will be extinguished, or will achieve fixation throughout a population. Within investigations of network-reciprocated cooperation~\cite{nowak_evolutionary_1992,nowak_five_2006}, models which abstract social networks to test mechanisms for the emergence of cooperation broadly follow approaches (implicitly) of this nature (see~\cite{perc_coevolutionary_2010} for a review of such investigations). 

The overwhelming majority of research studies in this field have considered a population to be one single connected network. However in the real world, multiple (relatively) discrete dynamic networks exist within populations, and at times, members of one social network may migrate to another. This is an aspect of cooperation in real-world scenarios which requires understanding, yet has thus far received little attention. In the work that follows, we isolate the key elements of such a scenario: namely, we have a primary network of interest, predicated on defector behaviour, and we consider the arrival of a very small group of connected individuals that have emigrated from a cooperative network.

Our investigations here also derive from a second motivating principle. In earlier work~\cite{miller_population_2015}, we have described how population size fluctuation has a positive impact, in promoting the emergence of cooperation in networks.  Commenting on this (ibid.), we suggested the possibility that the observed  effect may be viewed as a generalised response to perturbation of networks, and that population size fluctuation may be only one way, amongst several, of perturbing a network to thus yield similar results. This notion hints at a potential issue: that models of cooperation which are overly deterministic, or lacking in noise, may preclude the cooperative phenomena we seek to investigate. In the work that follows we consider whether our findings add further support to this thinking.

\section{Background}

Here we highlight a few key elements of game theory relevant to this work.  We then briefly consider existing research forming the basis for our investigations. 

Within the context of evolutionary game theory, a variety of games are used to model social behaviours.  A model of particular interest for investigating cooperation is the public goods game (PGG), otherwise referred to as the tragedy of the commons~\cite{hardin_tragedy_1968} or the n-person prisoner's dilemma.  This game, being based on group-wise rather than pair-wise behaviour, is arguably more analogous to the complexity of real-world social interactions, than the standard prisoner's dilemma (PD), which only models interactions between paired individuals~\cite{perc_evolutionary_2013}. 

In the PGG, each participant can choose to contribute, or not, a fixed amount to a central `pot'.  This pot is then increased by a multiplier and redistributed amongst all participants, regardless of whether they contributed.  The rational analysis of this game demonstrates that the selfish choice (defection), is the option which maximises an individual's payoff, however if all individuals exercise the same rationality, none will contribute and the public good will be minimised, hence we have a `social dilemma'. Whilst the rational analysis predicts tragedy, real-world examples of cooperation (contributing to the public good) are abundant.  It is this discrepancy between game theoretic predictions and empirical findings which research attempts to redress. 

The PGG can be implemented within \emph{evolving} social networks~\cite{santos_social_2008}, using an approach where each member of the network in turn, initiates a PGG within a group which consists of the individuals it is directly connected to---its `neighbourhood'. Any given individual in the network will be a neighbour of several other nodes, hence in addition to the PGG that a particular node initiates itself, it will also be a participant in PGGs initiated by others. It is this participation of an individual in multiple games with multiple opponents, i.e. \emph{group-wise} interaction, which differentiates the PGG from its cousin in game theory---the prisoner's dilemma (PD).  In the PD, an individual is able to retaliate or reciprocate in response to their partner's behaviour. In the PGG however, participants are not able to effectively target retaliation directly against defectors, since such retaliation (i.e. not contributing to the public good) harms cooperator and defector neighbours equally. The classical result for the PGG is that cooperation becomes less likely as neighbourhood size increases. This result can be appreciated intuitively, by considering that the more the neighbourhood size increases, (i.e. the closer it gets to having all members of the network participating), the more the game approximates the mean field scenario, where defection is the Nash equilibrium~\cite{nash_non-cooperative_1951}.   

The above approach has been extended to demonstrate the emergence of cooperation, amongst evolving populations of individuals playing PGG, in \emph{dynamic randomly growing} networks~\cite{miller_emergence_2016}. This development differs from earlier work in its use of two evolutionary elements, rather than one. The two elements are: 

\begin{enumerate}[nolistsep]
	\item \emph{Strategy updating}: This is the primary evolutionary mechanism, present in~\cite{santos_social_2008} and common to the majority of evolutionary game theoretic models used to investigate cooperation in networks. It represents intrinsic effects within the population, specifically, direct competition between two competing neighbours. This mechanism's effect is directly responsible for the spread of those strategies which confer greater fitness upon individuals. It does not however, in any way, affect the network topology.
	
	\item \emph{Population size fluctuation}: This secondary evolutionary mechanism~\cite{miller_minimal_2015} represents widespread `environmental' effects that are explicitly extrinsic to the population. In the real world, examples might be disease, predation, food shortages, drought, many of which may be seasonal. Here a proportion of the less fit members of a society are periodically `killed off'. Specifically, in the case of our implementation, individuals are removed from the population, along with the positions they occupied within the network due to their connections. This (fitness-based) process causes changes in the network topology, but it does not implement the spread of behaviours from one individual to another.   
\end{enumerate}

\noindent In the following, we investigate how a variety of network simulations, all predicated on originally non-cooperative behaviour, are affected by the arrival of a very small ($n \leq 3$) group of cooperative migrants.  We initially describe, in detail, the implementation of our models.  We then provide `behaviour profiles' for a range of network scenarios and growth mechanisms, followed by deeper scrutiny of phenomena within the actual simulations that are of particular interest.

\section{Methods}

Our work is based on methodology presented in~\cite{santos_social_2008,miller_emergence_2016,poncela_complex_2008}. We here give a full description of our approach for completeness.  
 
Our model describes agents located at the nodes of networks. Each node in the network has a `neighbourhood', defined by the nodes its edges connect to. A PGG occurs for each neighbourhood and hence a network of $N$ nodes will result in $N$ PGGs. Each agent in the network has a behaviour encoded by a `strategy' variable: `cooperate' or `defect', which determines whether it contributes to PGGs, or not, respectively. 

The general outline of the evolutionary process, for one generation, is as follows:

\begin{enumerate}[noitemsep]
	\item \emph{Play public goods games}: In a round-robin fashion, each agent initiates a PGG involving its neighbours.  An agent's fitness score is the sum of payoffs from all the individual PGGs that it participates in.
	\item \emph{Update strategies}: Selection occurs. Agents with low scores will have their strategies replaced, on a probabilistic basis, by comparison with the fitness scores of randomly selected neighbours.
	\item \emph{Remove nodes}: If the network has reached the nominal maximum size, it is pruned by a tournament selection process that removes less fit agents.
	\item \emph{Grow network}: A specified number of new nodes are added to the network, each connecting to $m$ randomly selected distinct existing nodes via $m$ edges. 

\end{enumerate}

\noindent In the following, we provide more detail on each of the four steps:

\smallskip

\noindent \textbf{\emph{Play public goods games}}. Each node of the network, in turn, initiates a PGG.  Within a single PGG, all cooperator members of a neighbourhood contribute a cost $c$ to `the pot'.  The resulting collective investment $I$ is multiplied by $r$, and $rI$ is then divided equally amongst all members of the neighbourhood, regardless of strategy.

Since an agent contributes a cost $c$ to each game they participate in, their overall contribution, in one generation, is therefore $c(k+1)$  where $k$ is the number of neighbours (degree). The single game individual payoffs of an agent $x$ are given by the following equations, for scenarios where $x$ is a defector ($P_D$) and a cooperator ($P_C$) respectively:

\begin{equation}
P_D = crn_c/(k_x+1) \enspace,
\label{eqn:FCPG_defector_scoring}
\end{equation}

\begin{equation}
P_C = P_D-c \enspace,
\label{eqn:FCPG_cooperator_scoring}
\end{equation}

\noindent where $c$ is the cost contributed by each cooperator, $r$ is the reward multiplier, $n_c$ is the number of cooperators in the neighbourhood based around $x$, and $k_x$ is the degree of $x$. 

\smallskip

\noindent \emph{\textbf{Update strategies}}. Each node $i$ selects a neighbour $j$ at random. If the fitness of node $i$, $f_i$ is greater or equal to the neighbour's fitness $f_j$, then $i$'s strategy is unchanged. If the fitness of node $i$, $f_i$ is less than the neighbour's fitness, $f_j$, then $i$'s strategy is replaced by a copy of the neighbour $j$'s strategy, according to a probability proportional to the difference between their fitness values. Thus poor scoring nodes have strategies displaced by those of more successful neighbours. 

Hence, at generation $t$, if $f_{i}(t)< f_{j}(t)$ then $i$'s strategy is replaced with that of the neighbour $j$ with the following probability:

\begin{equation}
	\Pi_{U_i}(t) = \frac
	{f_j(t) - f_i(t)}
	{fd\_max(k_i(t),k_j(t))} \enspace,
	\label{eqn:strategy_updating}
\end{equation}

\noindent where $k_{i}$ and $k_{j}$ are degrees of node $i$ and its neighbour $j$ respectively. The purpose of the denominator is to normalise the difference between the two nodes, with $fd\_max(k_{i}(t),k_{j}(t))$ representing the largest achievable fitness difference between the two nodes given their respective degrees. In the absence of a mathematical approach to calculate this, we run simulations for all 4 combinations (of the 2 strategy types at the 2 nodes), to establish maximum possible difference. 

\smallskip

\noindent \emph{\textbf{Grow network}}. We add 10 new nodes (7 on the first generation), with randomly allocated strategies, per generation. Each new node uses $m$ edges to connect to existing nodes.  Duplicate edges and self-edges are not allowed. The probability $\Pi(t)$ that an existing node $i$ receives one of the $m$ new edges is given by the following equations, for random attachment (RA), degree-based preferential attachment (PA), and fitness-based evolutionary preferential attachment (EPA)~\cite{poncela_complex_2008}, respectively:

\begin{equation}
	\Pi_{RA_i}(t) = \frac {1} {N(t)} \enspace,
	\label{eqn:RA_node_addition}
\end{equation}

\noindent where  $N(t)$ is the number of nodes available to connect to at time $t$ in the existing network.   (Given that in our model each new node extends $m = 2$ new edges, and multiple edges are not allowed, $N$ is therefore sampled \emph{without replacement}.) 

\begin{equation}
	\Pi_{PA_i}(t) = \frac
	{k_i(t)}
	{\sum_{j=1}^{N(t)}(k_j(t))} \enspace,
	\label{eqn:PA_node_addition}
\end{equation}

\noindent where $k_i(t)$ is the degree of an existing node \textit{i} and $N(t)$ is the number of nodes available to connect to at time \textit{t} in the existing network.  

\begin{equation}
	\Pi_{EPA_i}(t) = \frac
	{1 - \epsilon + \epsilon f_i(t)}
	{\sum_{j=1}^{N(t)}(1 - \epsilon + \epsilon f_j(t))} \enspace,
	\label{eqn:EPA_node_addition}
\end{equation}

\noindent where $f_i(t)$ is the fitness of an existing node \textit{i} and $N(t)$ is the number of nodes available to connect to at time \textit{t} in the existing network. The parameter $\epsilon \in [0,1)$ is used to adjust selection pressure. (We used $\epsilon = 0.99$ for `strong' EPA.)
 
Growth only occurs at times when the network is below a nominal maximum size (we used $N_{max} = 1000$ nodes). For all added nodes, other than migrants, we set $m=2$.

\noindent \emph{\textbf{Remove nodes (for fluctuation simulations)}}. Whenever the network achieves or exceeds the nominal maximum size, it is pruned by a percentage $X$. This is achieved by tournament selection using a tournament size equivalent to $1\%$ of the network. Tournament members are selected randomly from the network. The tournament member having the least fitness is the `winner' and is added to a short list of nodes to be deleted. Tournament selection continues until the short list of $X\%$ nodes for deletion is fully populated. 

The nodes on the short list (and all of their edges) are removed from the network. Any nodes that become isolated from the network as a result of this process are also deleted. (Failure to do this would result in small numbers of single, disconnected, non-playing nodes, having static strategies and zero fitness values.) When there are multiple nodes of equivalent low fitness value, oldest nodes are deleted first. Where $X = 0$, no deletions occur; in this case, on reaching maximum size, the network structure becomes static.\\

\noindent \textbf{Migrant clique attachment}. At generation 300, the migrant group connects to the existing primary network. Our migrant groups are small complete networks i.e. cliques, consisting of between 1 to 3 nodes (specific details in results section), all having cooperator strategies.  Initial connection to the primary network is via only one of the nodes in the clique. This node extends either 1 or 2 edges (specific details in results section) to existing network nodes chosen at random.  Once connected, the migrants are treated as a part of the primary network and are exposed to all elements of the evolutionary process described above. \\  

\noindent\textbf{General simulation conditions}. In networks grown from founder members, initial nodes were populated with defector strategies. In `pre-existing' networks, all nodes were populated with defectors. Strategy types of subsequently added nodes were allocated independently, uniformly at random (cooperators and defectors with equal probability). All networks had an overall average degree of approximately $k = 4$, hence an average neighbourhood size of $g = 5$ (since neighbourhood includes self). Simulations were run until 20,000 generations. Final `fraction of cooperators' values we use are means, averaged over the last 20 generations of each simulation. Each simulation consisted of 25 replicates. We used a shrinkage value of $X$ = 2.5\% for all fluctuation simulations. Simulation data is recorded after step 2 (\emph{Update strategies}).

\section{Results and Discussion}

\begin{figure}[!h]
	\begin{center}
		\includegraphics[width=12.2cm]{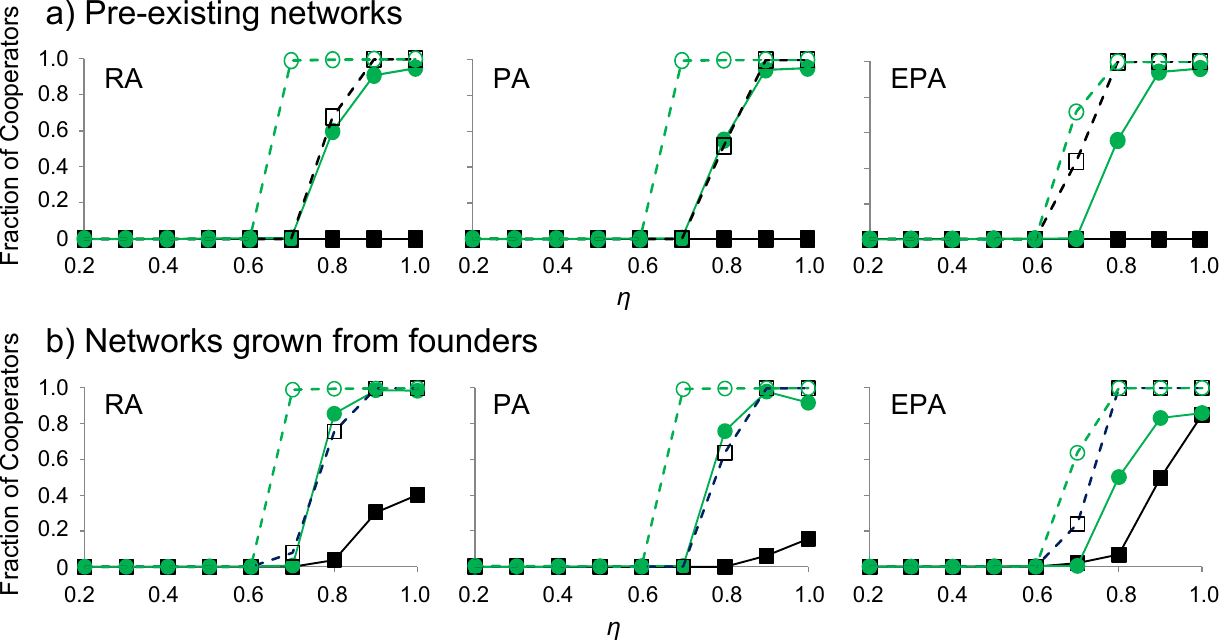}
		\caption{Behaviour profile plots illustrating the impact of a migrant cooperator clique on the emergence of cooperation for three different attachment mechanisms: RA, PA and EPA. \textbf{a}) shows pre-existing networks having initially random graph topology and initialised entirely with defectors. \textbf{b}) shows networks grown from 3 defector founders. Final fraction of cooperators present is plotted against $\eta$ (the PGG reward multiplier $r$ normalised with respect to average neighbourhood size, $g = 5$). Migrant cliques consist of 3 connected cooperators, one of which attaches to the existing network randomly by 2 nodes. Green lines with circle markers are simulations featuring migrants. Black lines with square markers are controls (no migrants). Solid lines represent simulations that are fixed in their network topology. (In the case of \textbf{b}, topology becomes fixed upon population achieving maximum size.) Dashed lines represent fluctuating simulations.}
		\label{fig:Cooperation_profiles}
	\end{center}
\end{figure}

We initially present our results using an approach common for investigations in this field. We aggregate data from multiple differing sets of simulations, plotting final fraction of cooperators against the variable, $\eta$, which is the PGG reward multiplier normalised with respect to the average neighbourhood size in the network. In Fig. \ref{fig:Cooperation_profiles}a we present such `behaviour profiles' for results from the `simplified scenario' of pre-existing networks. These networks have initially random graph topology~\cite{erdos_random_1959} and are initialised entirely with defectors. Figure \ref{fig:Cooperation_profiles}b, illustrates the `more realistic' scenario where we consider networks grown from their origins, in this case from 3 founder defector members. In both network scenarios we provide profiles for the three attachment mechanisms of RA, PA and EPA.

For the simpler scenario of pre-existing networks, initialised with all defectors (Fig. \ref{fig:Cooperation_profiles}a) and having a fixed network size, we naturally observe zero cooperation (solid black lines) for all attachment mechanisms.  In comparison, the migrant scenario (solid green line) precipitates cooperation once the temptation of the reward achieves a particular threshold ($\eta > 0.7$).  In the case of fluctuating network size, we see that the migrants promote higher levels of cooperation than those seen in their absence (compare green dashed with black dashed lines), except in the case of EPA, where levels of cooperation have already been elevated by the increased network heterogeneity associated with this mechanism (see~\cite{santos_new_2006} for detailed information on the role of network heterogeneity in cooperation).

When we consider the more complex scenario of networks grown from founders (Fig. \ref{fig:Cooperation_profiles}b), we see that our earlier findings still hold.  Again, above a reward threshold ($\eta > 0.6$), the arrival of the migrants promotes widespread cooperation.  We see this effect for networks that become static on reaching specified maximum size and also in those that fluctuate in size thereafter. We note that in the case of fluctuating models, we see little difference in final outcomes when comparing pre-existing networks with those grown from founders (compare corresponding coloured dashed lines in Figs. \ref{fig:Cooperation_profiles}a and b). As described in earlier research~\cite{miller_minimal_2015}, the fluctuation mechanism, by deleting low fitness nodes from within the network, can overcome the limitations of `fossilised' (zero-fitness, defector-dominated) regions of the network in a manner that is not achievable by strategy updating between neighbours. Importantly, cooperation can be supported by a fluctuating population size, without the requirement for highly heterogeneous network topology: the fluctuation mechanism drives networks to a topology that has only moderately heterogeneous connectivity (in the form of a compressed exponential degree distribution)~\cite{miller_emergence_2016}.	
 
Whilst the behaviour profile plots above allow us to neatly characterise and compare different experimental simulations, they describe derived data which for the most part is of limited interest, whilst potentially masking more interesting phenomena. More specifically, as the value of the reward variable ($\eta$) is maximised/minimised, the dilemma becomes diminished and the dominant behaviour of populations becomes consistent and highly predictable. We suggest that in presenting abstracted representations of real-world scenarios, such regions of the behaviour profiles are of limited relevance.

It is the \emph{mid-range} values of the reward variable that represent the social dilemma in its strongest form.  We suggest that these regions are of particular importance in investigating the emergence of cooperation, since they represent the much more realistic challenge faced in nature by individuals attempting to \emph{balance} cost versus reward, and in addition, where noise may likely be a confounding or contributory factor.  Where we see transitions in population behaviour, where a mixture of competing behaviours exists, where the choice of cooperate or defect is not clear cut, and where noise may be present---these are the areas we are interested in.

\begin{figure}[!h]
	\begin{center}
		\includegraphics[width=12.2cm]{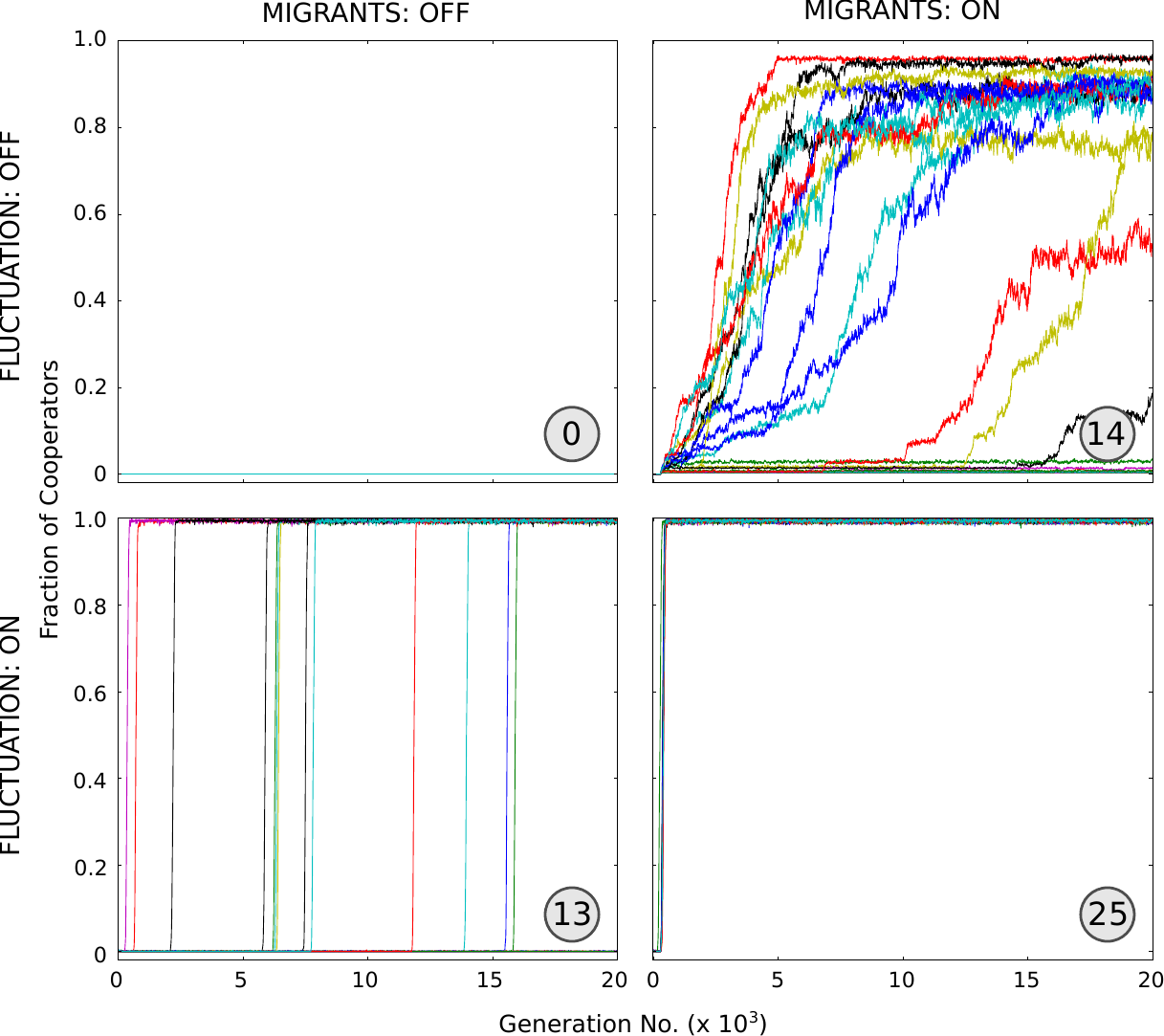}
		\caption{Simulation time plots (25 replicates) illustrating the effects of migrant clique arrival and fluctuation, in pre-existing random networks initialised with defectors, with $\eta = 0.8$. Plots show number of cooperators over 20,000 generations. Migrant groups are complete networks of 3 cooperator nodes, 1 of which connects to 2 randomly selected existing network nodes. Network growth is by random attachment. All other details are as described in Methods section. Number of replicates transitioned to cooperation is shown in circle inset.}
		\label{fig:Time_plots_PENs}
	\end{center}
\end{figure}

We now explore the behaviour of our populations, in these regions of interest, by focusing on the behaviour of replicate simulations as they transition from defection to cooperation.  From Figs. \ref{fig:Cooperation_profiles}a and b, we see the widest variety of outcomes in the region approximately around where $\eta = 0.8$. Figure \ref{fig:Time_plots_PENs} illustrates individual time plots of simulations based on this value, for the simplified case of pre-existing networks initialised with defectors.  The plots show simulations with the effects of fluctuation and immigration enabled, disabled, and acting in concert.  We summarise from inspection of these plots that:

\begin{enumerate}[label=\roman*, nolistsep]
	\item The fluctuation mechanism on its own enables a majority of replicates to transition to cooperation.  Similar levels of cooperation are achieved by all of those replicates that transition. Transition times however remain variable with some replicates failing to transition over the time period studied.

	\item The isolated effect of migrant arrival drives higher levels of cooperation amongst replicates.  In the case of this effect though (in contrast to our previous observation), it is the \emph{levels} of cooperation achieved which are variable.
		
	\item The combined impact of migrants together with fluctuating population size, results in all replicates transitioning to cooperation with consistency in both final levels of cooperation achieved, and also in transition times (all replicates transition within 200 generations of the arrival of the migrants). 
\end{enumerate}

\begin{figure}[!h]
	\begin{center}
		\includegraphics[width=12.2cm]{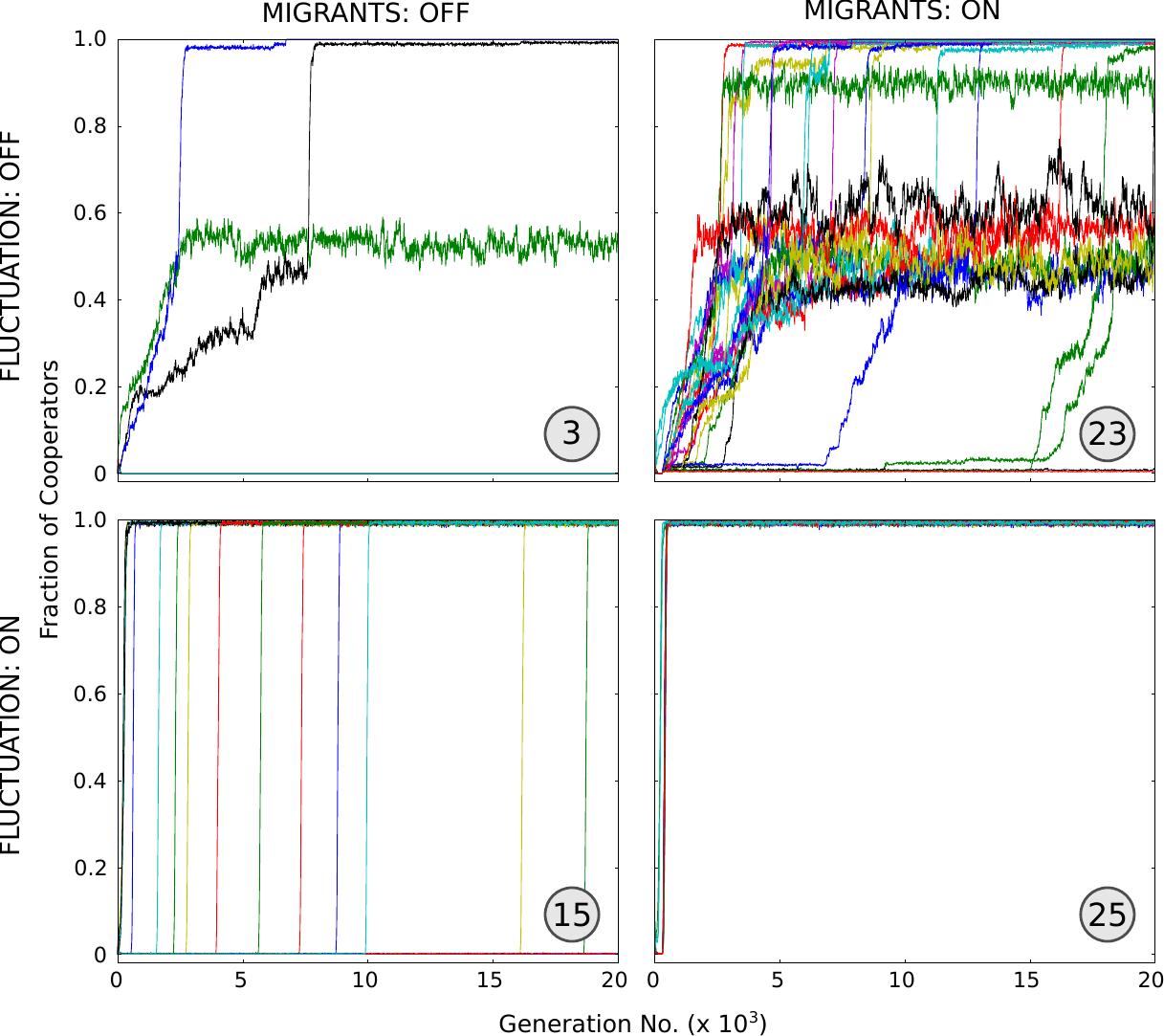}
		\caption{Simulation time plots (25 replicates) illustrating the effects of migrant clique arrival and fluctuation, for networks grown from 3 defector founders, with $\eta = 0.8$.  Plots show number of cooperators over 20,000 generations. Migrant groups are complete networks of 3 cooperators nodes, 1 of which connects to 2 randomly selected existing network nodes. Network growth is by random attachment. All other details are as described in Methods section. Number of replicates transitioned to cooperation is shown in circle inset.}
		\label{fig:Time_plots_seeds}
	\end{center}
\end{figure}

In Fig. \ref{fig:Time_plots_seeds} we illustrate similar time plots, in this case for the more complex scenario featuring networks grown from founder populations of 3 defectors.  We observe that the findings seen earlier, for the simplified case of pre-existing networks, still hold: fluctuation alone promotes consistent levels of increased cooperation albeit with variable transition times; migrants alone promote cooperation albeit to varying levels; the combination of cooperator migrants and fluctuation brings consistency to both transition times and levels of cooperation achieved.

These findings are also robust to attachment mechanisms.  For both of the network models illustrated above, in addition to random attachment (as represented in Figs. \ref{fig:Time_plots_PENs} and \ref{fig:Time_plots_seeds}), the same observations also held when tested using both degree-based attachment (PA), and fitness-based attachment (EPA).

The ability of the of the migrant clique to invade defector networks appears to arise from benefits conferred on the connecting migrant by the `back-up' provided from its fellow migrants.  These back-up migrants are initially immune to both strategy updating and the impact of defectors in reducing their payoff values (being as they are initially not directly connected to the network). The back-up migrants can boost the payoff (fitness) of a connecting migrant, so that during strategy updating, it can thus readily convert the existing network node it connects to, into a cooperator. Beyond initial possible payoff calculations, which can be established analytically, it becomes harder to pin down the details of the further spread of cooperation. However, it is clear from our investigations that in the case of migrant-triggered cooperation, it is this back-up which is key.  

What is particularly interesting here, is just how small the migrant group can be, whilst still being able to precipitate the emergence of cooperation through the entire population.  The previous simulations were based on migrant groups of 3 connected individuals, one of which extends 2 connections to random existing members of the network.  In additional work, we have reduced the size of the migrant group to 2 individuals, of which one connects only 1 edge to an existing network node. Tested at the same $\eta$ ($= 0.8$), on pre-existing defector-populated initially random networks, and on networks grown from defector founders (growth by RA in both cases), our previous findings still hold. (Time plots were highly similar to those shown in Figures 2 and 3, with the only difference that a delay in transition was observed infrequently, e.g. 1 or 2 replicates out of 25, for those simulations combining both migration and fluctuation.)  On further reduction to 1 node (extending either 1 or 2 edges), our general findings no longer hold.  This outcome is entirely expected, as this situation is now no different to the standard attachment process by which all new individuals routinely connect---1 node, 2 edges, i.e. no back-up.

These findings based on adjustments to the migrant clique highlight a potential source of concern regarding models of cooperation in networks, namely that widely differing outcomes may arise from seemingly small differences in simulation parameters: We can reduce our migrant mechanism to a point where it appears very similar (2 nodes, 1 edge) to the mechanism by which nodes routinely attach during network growth (1 node, 2 edges). Given such similarity, and noting that the migrant effect happens only once in a simulation, whilst new nodes are added repeatedly in the fluctuation model, we might be inclined to therefore assume that results due to the migrant clique arrival would be trivial relative to those arising from fluctuation. However, we see in our results that the isolated, seemingly trivial, migrant event clearly brings about an additional change to populations, which is not achieved in its absence. The small difference between these two very similar mechanisms results in markedly different behavioural dynamics. Importantly, despite their apparent similarities, the attachment mechanism used for routine network growth clearly cannot create the additional opportunities for cooperation that the migrant clique's arrival can enable. 

These results combined with findings of previous research, reinforce our belief that fluctuations in the network, or migrant cliques, or alternative mechanisms to perturb the system, bring an added dimension to models of cooperation in networks that simpler mechanisms fail to provide: It is these noisy perturbations of the network that disrupt the `status quo' and catalyse the spread of cooperation throughout the population.  If this assumption is correct then there is a risk that simpler, more deterministic models of cooperation in networks may lack the disruptive elements that promote cooperation and may thus preclude or impede its emergence.

\section{Conclusion}

Using various models of cooperation, based on the public goods game, we have investigated a scenario where individuals migrate, from a cooperative network, to join one that does not demonstrate cooperation.  Under certain conditions, notably around the region where the social dilemma is at its strongest, we find quite striking results: The effect of a few concerted migrants catalyses a marked behavioural change, precipitating the widespread emergence of cooperation throughout the entire population. Of particular interest is our finding that the migrant group size can be extremely small and needs only to form one initial connection in order to initiate a marked response. The actions of a seemingly trivial group of concerted cooperators initiate changes throughout a population that is orders of magnitude larger than the migrant group. 

We have hypothesised that perturbation, in the form of population size fluctuation, and also in the form of invading migrants, can promote cooperation. We have demonstrated this to be the case for both of these effects in isolation, and to a greater extent, in concert. Clearly other methods, or combinations of methods, for perturbing or disrupting networks exist that may yield similarly interesting results. 

Our results reinforce previous work proposing that perturbations of networks, or possible alternative forms of disruption, are an important contributory feature in the emergence of cooperation. Taken generally, such observations suggest the potential for oversimplified or strictly deterministic models of cooperation in social networks, to limit or exclude the phenomena they seek to investigate. We highlight, in particular, that from a combination of two mechanisms studied here, there emerged a consistency in outcome that is unlikely to have been anticipated from studying simpler models of each mechanism in isolation.

\subsubsection*{Acknowledgements}

This work has been funded by the Engineering and Physical Sciences Research Council (Grant reference number EP/I028099/1).

\bibliographystyle{splncs}
\bibliography{SGM_UCNC_references}

\end {document}